\documentclass[12pt]{iopart}

\usepackage[utf8]{inputenc}
\usepackage[numbers]{natbib}
\usepackage{graphicx}
\usepackage{amssymb}
\usepackage{dcolumn}
\usepackage{bm}
\usepackage{color}
\usepackage[utf8]{inputenc}
\usepackage{capt-of}
\usepackage{tabu}
\usepackage{mwe} 
\usepackage{subcaption}
\usepackage{soul}
\usepackage{xr}

\begin{document}

\title{Contagion dynamics in self-organized systems of self-propelled agents}

\author{Yinong Zhao$^{1,2}$, Cristi\'an Huepe$^{3,4,5}$ and Pawel Romanczuk$^{1,2}$}

\address{$^1$Institute for Theoretical Biology, Department of Biology, Humboldt-Universität zu Berlin, 10115 Berlin, Germany}
\address{$^2$Bernstein Center for Computational Neuroscience Berlin, 10115 Berlin, Germany}
\address{$^3$School of Systems Science, Beijing Normal University, Beijing 100875, China}
\address{$^4$CHuepe Labs, 2713 W Haddon Ave \#1, Chicago, IL 60622, USA}
\address{$^5$Northwestern Institute on Complex Systems and ESAM, Northwestern University, Evanston, IL 60208, USA}

\begin{abstract}
    We investigate the Susceptible-Infectious-Recovered contagion dynamics in a system of self-propelled particles with polar alignment.
    Using agent-based simulations, we analyze the outbreak process for different combinations of the spatial parameters (alignment strength and Peclet number) and epidemic parameters (infection-lifetime transmissibility and duration of the individual infectious period). 
    We show that the emerging spatial features strongly affect the contagion process.
    The ordered homogeneous states greatly disfavor infection spreading, due to their limited mixing, only achieving large outbreaks for high values of the individual infectious duration.
    The disordered homogeneous states also present low contagion capabilities, requiring relatively high values of both epidemic parameters to reach significant spreading. 
    Instead, the inhomogeneous ordered states display high outbreak levels for a broad range of parameters. The formation of bands and clusters in these states favor infection propagation through a combination of processes that develop inside and outside of these structures.
    Our results highlight the importance of self-organized spatiotemporal features in a variety of contagion processes that can describe epidemics or other propagation dynamics, thus suggesting new approaches for understanding, predicting, and controlling their spreading in a variety of self-organized biological systems, ranging from bacterial swarms to animal groups and human crowds.
\end{abstract}

\section{Introduction}

Contagion dynamics are ubiquitous in biological and social systems. They can encompass the spread of disease \cite{pastor2001epidemic,keeling2011modeling}, rumors \cite{nekovee2007theory,liu2014analysis,zhao2013sir}, behaviors \cite{rosenthal2015revealing,sosna2019individual}, or information \cite{gump1997stress,behnke1994contagion}.
These different types of phenomena are often modeled using generic descriptions that were developed to represent infection spreading and epidemic outbreaks, in which individuals can be in a susceptible, infectious, or recovered state.
In these models, the contagion typically propagates through an interaction network that plays a decisive role in the spreading dynamics.
In realistic scenarios, the interactions are time-dependent and determined by a proximity network that changes as a function of the positions of the individuals.
In cases where these agents follow complex collective dynamics, the changing topology can depend in turn on their self-organized motion.
To improve our modeling of real-world systems, it is therefore important to understand the relationship between the emergence of large-scale contagion processes and the evolving interaction networks that develop between agents that display collective behavior \cite{balcan2009multiscale}.

In spatially explicit models of mobile agents, in contrast to the idealized mean-field or network models, the potential contagion interactions are determined by a switching proximity network that changes in time as the individuals follow a specific set of spatial dynamical rules (for an example, see \cite{rahmani2020flocking}).
The core parameters of a typical mean-field SIR-contagion process, such as the contact rate and contact duration, then become emergent features of the self-organized spatial dynamics, as it was shown for gas-like systems of self-propelled active Brownian particles in \cite{gonzalez2004scaling,peruani2008dynamics,peruani2019reaction,norambuena2020understanding}.
The properties of the spatial dynamics, such as the speed of the individual agents considered in \cite{peruani2019reaction}, can thus affect in nontrivial ways the global contagion process.
In these examples, however, the agents or particles do not form spatial structures, displaying instead homogeneous density and isotropic orientations, in contrast to what is known to occur in many models of self-propelled agents and in real-world systems.
In particular, in models where spatial interactions such as velocity alignment forces between neighboring agents are introduced, we expect the formation of states that display long-range order (leading to coordinated collective movement at the scale of the system) and a variety of spatial structures (such as large-scale, high density bands or clusters) \cite{vicsek1995novel,chate2008collective,martin2018collective,caussin2014emergent}.
To properly describe the contagion process, we must therefore consider the various degrees of mixing and density distributions exhibited by different self-organized states.

In this work, we implement a Susceptible-Infectious-Recovered (SIR) infection process \cite{kermack1991contributions,keeling2011modeling}
in a system of self-propelled agents with alignment and repulsion interactions to explore the impact of self-organized spatiotemporal structures on the contagion dynamics.
Our results show that the epidemic outcome depends strongly on the specific emergent properties of each collective state.
We observe that the outbreak thresholds are significantly lower in ordered states with large-scale clusters or band-like structures, and higher in ordered or disordered states with homogeneous density distributions.
By comparing the contagion dynamics of different states, we find that the timescales of the emerging spatial structures and of the infection processes play an essential role in the epidemic outcome.

\section{Model}

We consider a system of $N$ self-propelled agents moving continuously in a two-dimensional arena of size $L\times L$, with periodic boundary conditions. At time $t$, agent $i$ is located at position $\vec{r}_i(t)$ and tends to advance in its heading direction 
$\hat n_i=\left[\cos\theta_i(t),\sin\theta_i(t) \right]^T$ with constant self-propulsion speed $v_0$, while also being displaced by local soft-core repulsion forces $\vec{F}_i$ between neighbors. The corresponding equation for a system with overdamped dynamics is given by
\begin{equation}
    \frac{d\vec{r}_i(t)}{dt} = v_0\hat{n}_i(t) + \vec{F}_i(t).
\end{equation}
We define the interaction set 
$S_i=\left\{j\,|\, |\vec{r}_{ji}|\leq \mathrm{r_{int}} \right\}$, with $\vec{r}_{ji}=\vec{r}_i-\vec{r}_j$,
as containing all neighbors within a range
$\mathrm{r_{int}}$ of the focal particle $i$.
In order to reduce the parameter space, the alignment and repulsive interactions are defined as having the same range. These interactions are given by
\begin{eqnarray}
    \frac{d\theta_i(t)}{dt} = \frac{1}{\tau}\left\langle \mathrm{mod}^{*}(\theta_j -\theta_i) \right\rangle_{j\in S_i} + \sigma\xi_{\theta} \\
    \vec{F}_i(t) = \sum_{j \in S_i}\frac{\mathrm{r_{int}}-|\vec{r}_{ji}|}{\mathrm{r_{int}}}\frac{\vec{r}_{ji}}{|\vec{r}_{ji}|}.
\end{eqnarray}
Here, $\tau$ is a relaxation time that controls the strength of the alignment,  $\mathrm{mod}^{*}(x)=\mathrm{mod}(x+\pi,2\pi)-\pi$ is a
modified modulo function that produces values between $-\pi$ and $\pi$, and $\sigma$ is the strength of a $\delta$-correlated Gaussian white noise introduced through a random variable $\xi_{\theta}$ that satisfies 
$\langle \xi_\theta \rangle=0$ and
$\langle \xi_{\theta}(t_1)\xi_{\theta}(t_2) \rangle = \delta(t_2 - t_1)$.
Each particle thus tends to align its heading angle $\theta_i(t)$ to its neighbors, while subjected to noise, and is displaced by repulsive forces $\vec{F}_i(t)$ that do not affect its heading direction.

In addition to the spatial dynamics, we implement a simple contagion process that does not affect the agent motion, corresponding to the standard SIR model in epidemiology.
Each agent $i$ is assigned an internal ternary variable $s_i(t) \in \{S,I,R\}$, representing its current contagion state: susceptible ($s_i(t)=S$), infected ($s_i(t)=I$), or recovered ($s_i(t)=R$).
A susceptible agent $i$ can only become infected, with a base transmission rate $\beta_b$, while it is in contact with at least one infected agent $j$ (i.e., at a distance smaller than the interaction range $|\vec{r}_{ij}|\leq\mathrm{r_{int}}$).
Its infection probability when in contact with a single infected neighbor for a short time period $\Delta t$ will thus be given by $p_{inf}=\beta_b\Delta t$. 
If a susceptible agent is in contact with multiple infected agents simultaneously, all pairwise interactions are assumed to be statistically independent and its infection probability is therefore given by the sum of the individual probabilities, until it saturates at $1$. 
On the other hand, an infected agent spontaneously recovers at a rate $\gamma$, that is, with probability equal to $\gamma \Delta t$ for a small enough $\Delta t$.
We consider here the recovered state to be an absorbing state of the contagion dynamics, so no infections can reoccur in recovered agents.

Note, that the base transmission rate $\beta_b$, as defined above, is different from the transmission rate parameter in standard epidemiological compartment models, typically denoted as $\beta$. For a short contact duration $\Delta t$ the following relations holds $\beta=\beta_b \nu \Delta t$, with $\nu$ being the contact rate.  

Following the SIR dynamics defined above, an initial number of infected agents can result in a contagion process that spreads to other agents, all of which will eventually recover spontaneously until no infected agents remain in the system, ending the outbreak.
If we define $\rho_I(t)$, $\rho_S(t)$ and $\rho_R(t)$ as the fraction of the total population that is infected, susceptible, or recovered at time $t$, respectively, this corresponds to reaching a state with 
$\rho_I(\infty)=0$ and $\rho_S(\infty)+\rho_R(\infty)=1$,
where $t=\infty$ denotes any time $t$ after the SIR dynamics has reached its final absorbing state. The effectiveness of the epidemic outbreak can then be measured by the value of $\rho_R(\infty)$.

We will investigate the contagion process for a system of self-propelled agents in different types of collective states \cite{zhao2021phase}. These correspond to distinct regions of the phase space that is spanned by two nondimensional control parameters, the Peclet number Pe and the effective coupling strength g, defined as
\begin{eqnarray}
    \mathrm{Pe} &=& \frac{v_0}{\mathrm{r_{int}}\sigma^2} \\
    \mathrm{g} &=& \frac{1}{\tau\sigma^2}.
\end{eqnarray}
The Peclet number quantifies the level of persistence in individual movement, with $\mathrm{Pe}\to 0$ corresponding to noise dominated Brownian motion and $\mathrm{Pe}\to\infty$ describing ballistic motion for vanishing angular noise. The effective coupling $\mathrm{g}$, quantifies the strength of the alignment interaction relative to the angular noise.

Each region in the (Pe,g) parameter space  is characterized by different degrees of alignment and clustering, which is measured by two order parameter, $\Psi$ and $\Lambda$, respectively.
More specifically, the degree of alignment is given by the polarization
\begin{equation}
    \label{eq:polarization}
    \Phi=\frac{1}{N} \left|\sum_{i = 1}^{N} \hat{n}_i\right|,
\end{equation}
where $\Phi=1$ if all particles are heading in exactly the same direction and $\Phi\approx 0$ if they are moving in random directions.
The clustering $\Lambda$ is defined as the fraction of the total number of particles that conform the largest cluster in the system, where the members of a cluster are determined to be all particles within a distance $\mathrm{r_{int}}$ of any other particle also in the cluster.
Therefore, for the relatively low mean density value considered in this paper, $\Lambda=1/N \ll 1$ describes a homogeneous state in which the distance between all neighbors is larger than the interaction range, whereas $\Lambda=1$ corresponds to a highly inhomogeneous state where all agents belong to a single cluster.

\section{Results}

We studied the contagion process of the SIR model in a system of self-propelled particles that follow the dynamics described above.
We considered different collective states in the (Pe,g) phase space by varying the noise $\sigma$ and alignment relaxation time $\tau$, while keeping a fixed preferred agent speed $v_0=0.2$ and interaction range $\mathrm{r_{int}}=1$, in order to limit the parameter exploration. 
We also fixed the mean density to a packing fraction of
$\mathrm{N} \pi \, {\mathrm{r}}_{\mathrm{int}}^2 / (4 \mathrm{L}^2) = 0.3$, by simulating 
$\mathrm{N}=10000$ agents in an arena of size $\mathrm{L}\approx 161$.
A detailed analysis of the different spatial dynamics and collective states obtained in this system was recently presented in \cite{zhao2021phase}. 
Starting from statistically stationary states of the spatial dynamics, we implemented the SIR process by infecting 100 agents, which corresponds to 1 percent of all particles. 
For each parameter combination, we considered 5 independently converged stationary states and initialized 10 contagion processes, each starting from a different set of randomly selected infected agents. Hence, a set of 50 simulations was used to characterize the epidemic spreading for each case.

\begin{figure}[t]
    \centering
    \begin{tabular}{cc}
    \begin{subfigure}[b]{0.4\textwidth}
        \includegraphics[width=1.1\textwidth]{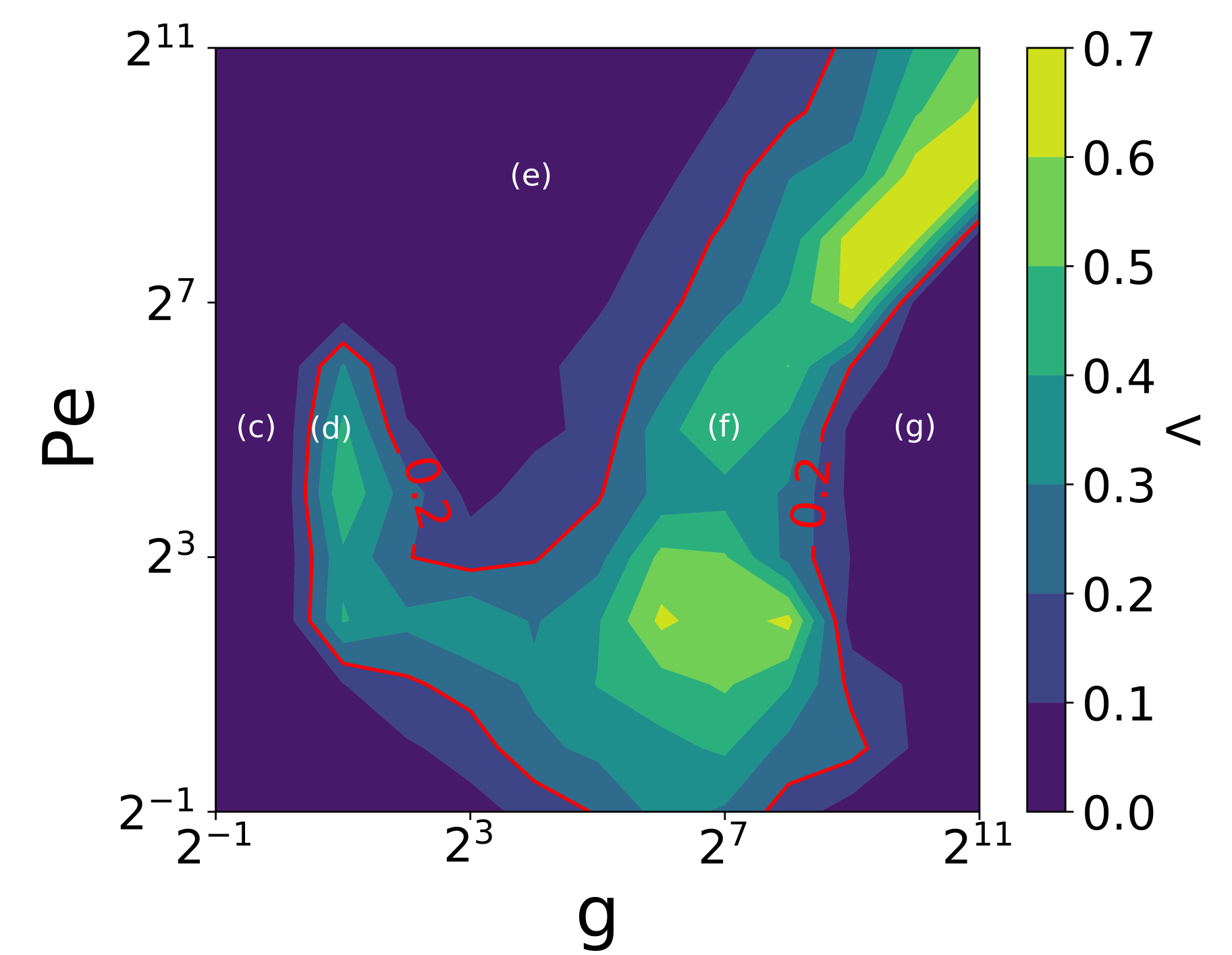}
        \caption{Largest cluster fraction}
        \label{fig:clustering-diagram}
    \end{subfigure}
    \hspace{0.5cm}
    \begin{subfigure}[b]{0.4\textwidth}
        \includegraphics[width=1.1\textwidth]{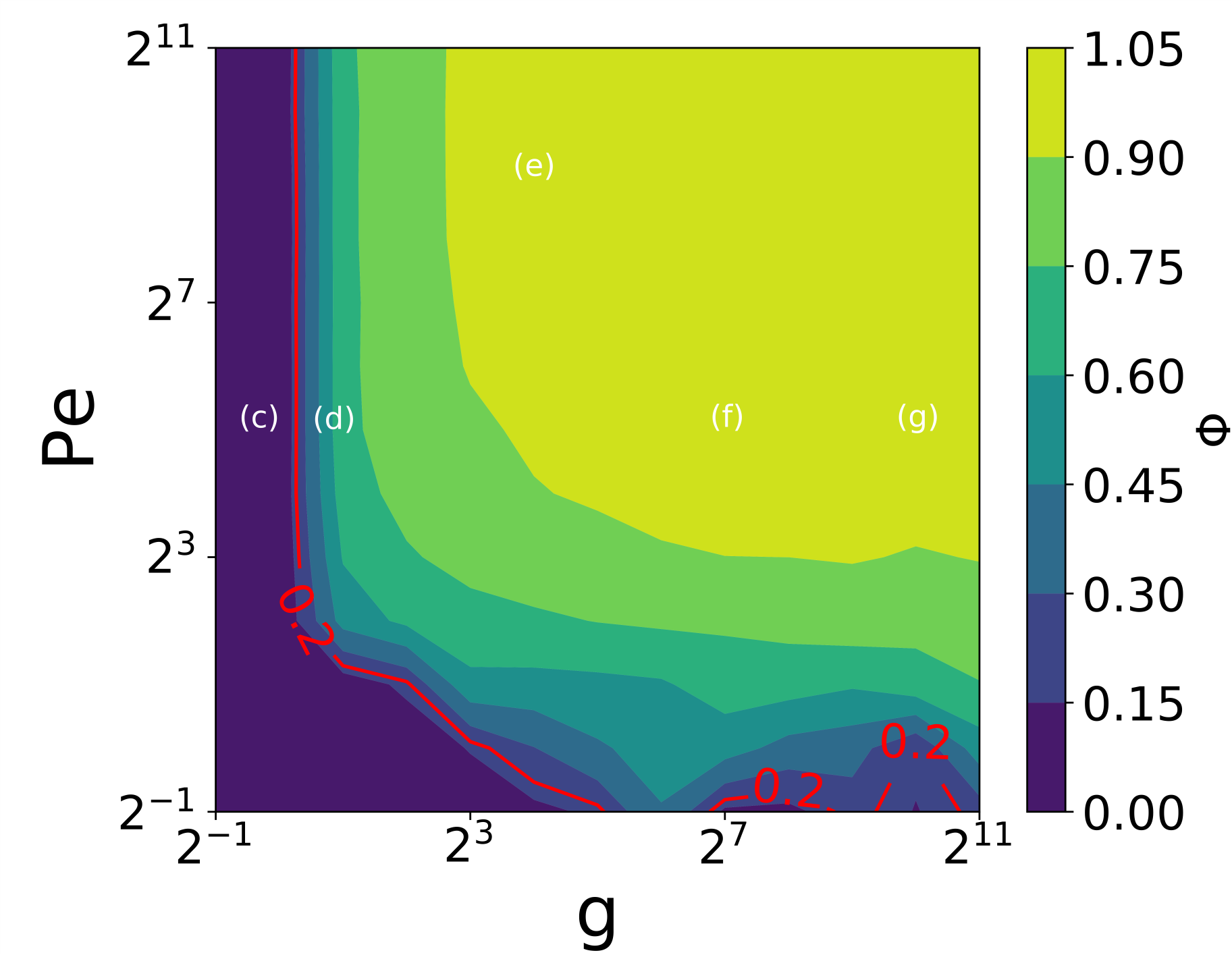}
        \caption{Polarization}
        \label{fig:polarization-diagram}
    \end{subfigure}
    \end{tabular}
    
    \begin{tabular}{ccccc}
        \begin{subfigure}[b]{0.16\textwidth}
            \includegraphics[width=1.2\textwidth]{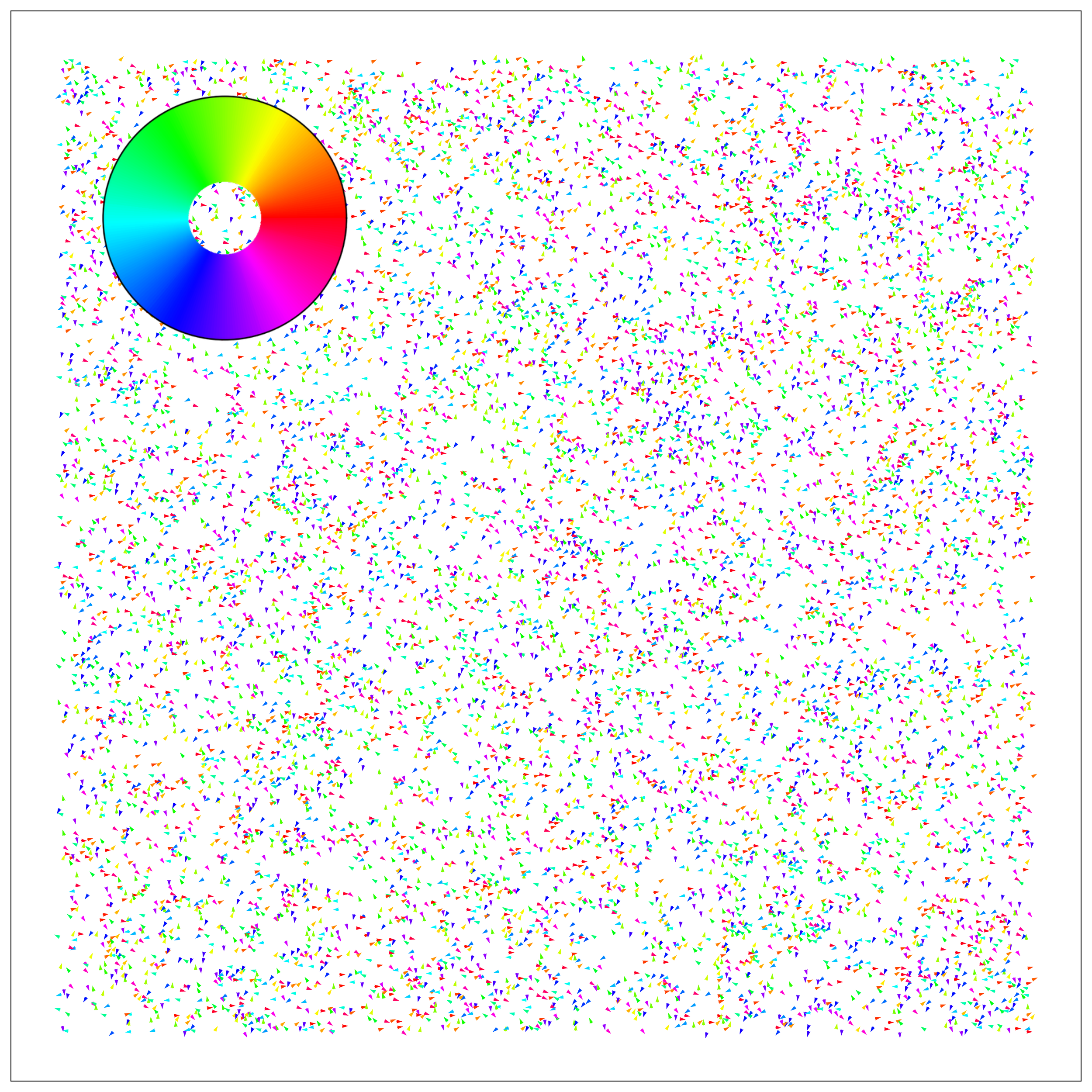}
            \caption{Disordered}
        \end{subfigure}
    &
        \begin{subfigure}[b]{0.16\textwidth}
            \includegraphics[width=1.2\textwidth]{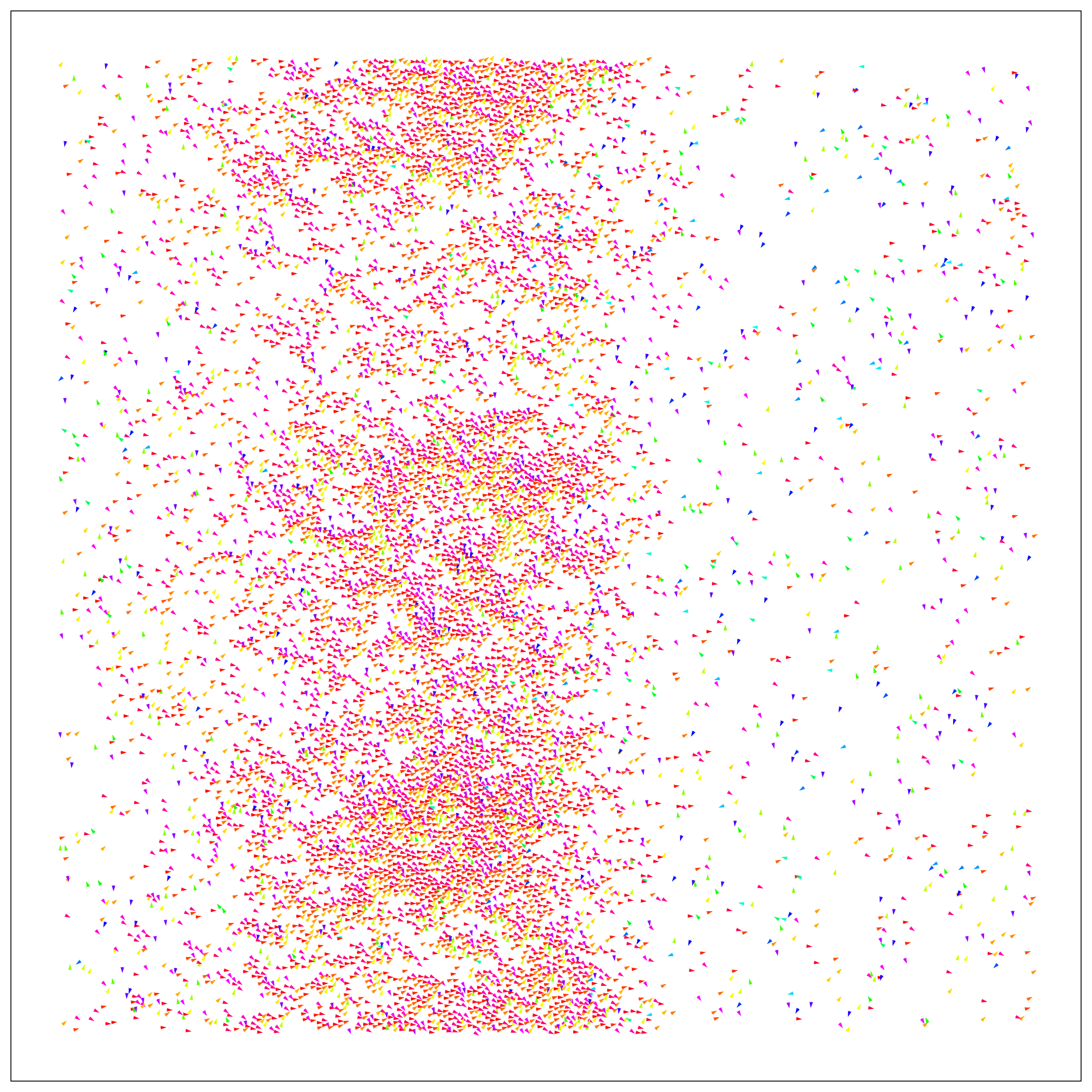}
            \caption{OB}
        \end{subfigure}
    &
        \begin{subfigure}[b]{0.16\textwidth}
            \includegraphics[width=1.2\textwidth]{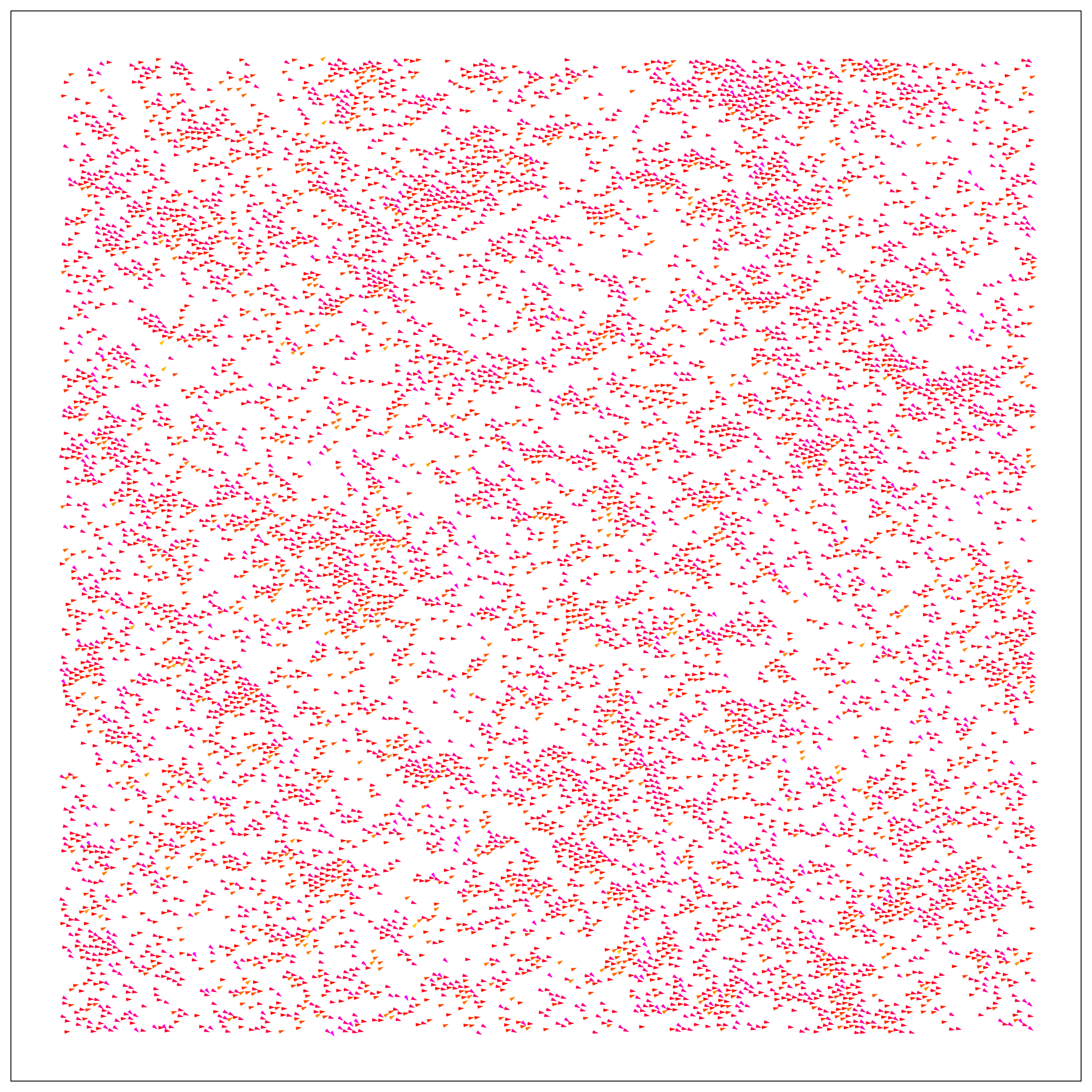}
            \caption{OH-1}
        \end{subfigure}
    &
        \begin{subfigure}[b]{0.16\textwidth}
            \includegraphics[width=1.2\textwidth]{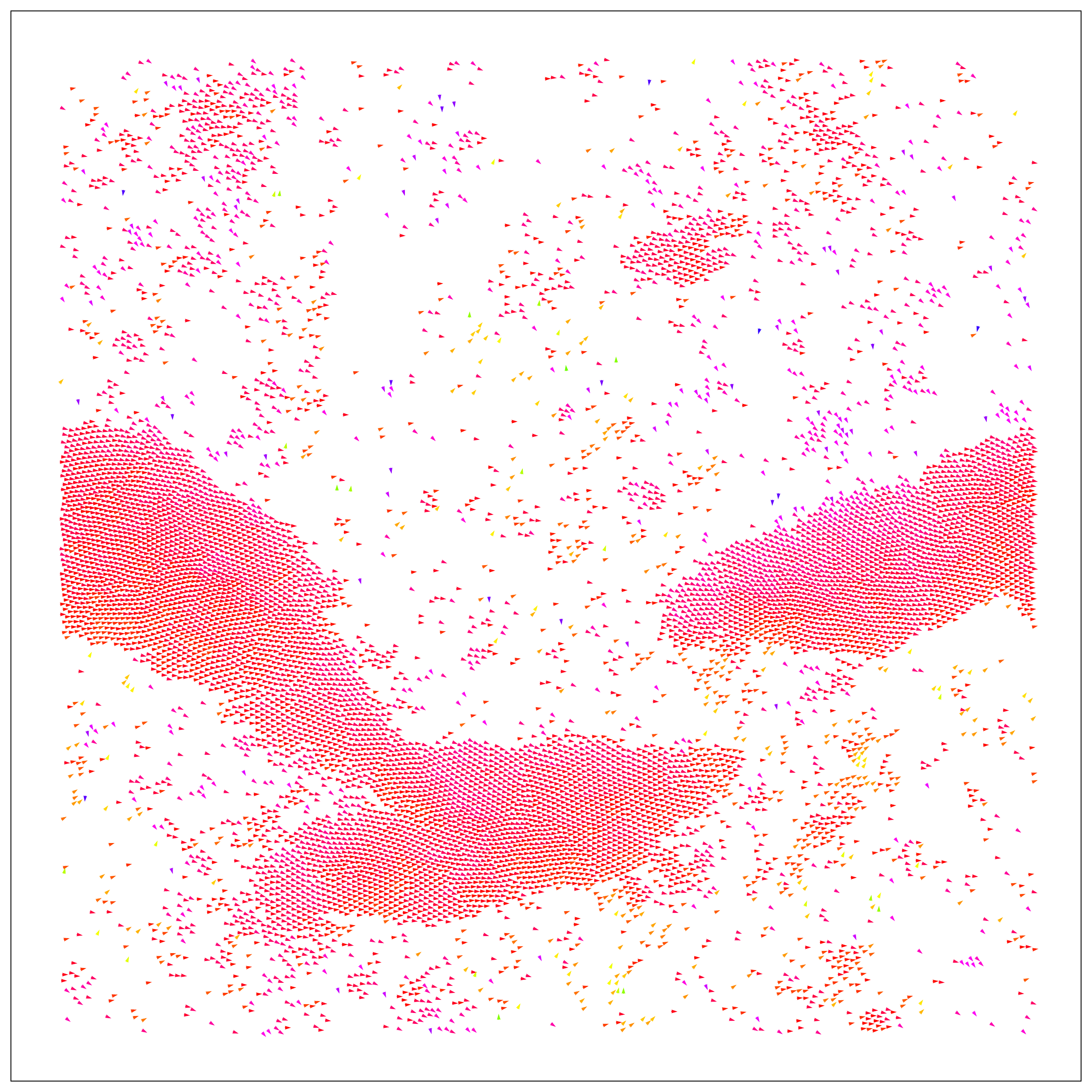}
            \caption{OC}
        \end{subfigure}
    &
        \begin{subfigure}[b]{0.16\textwidth}
            \includegraphics[width=1.2\textwidth]{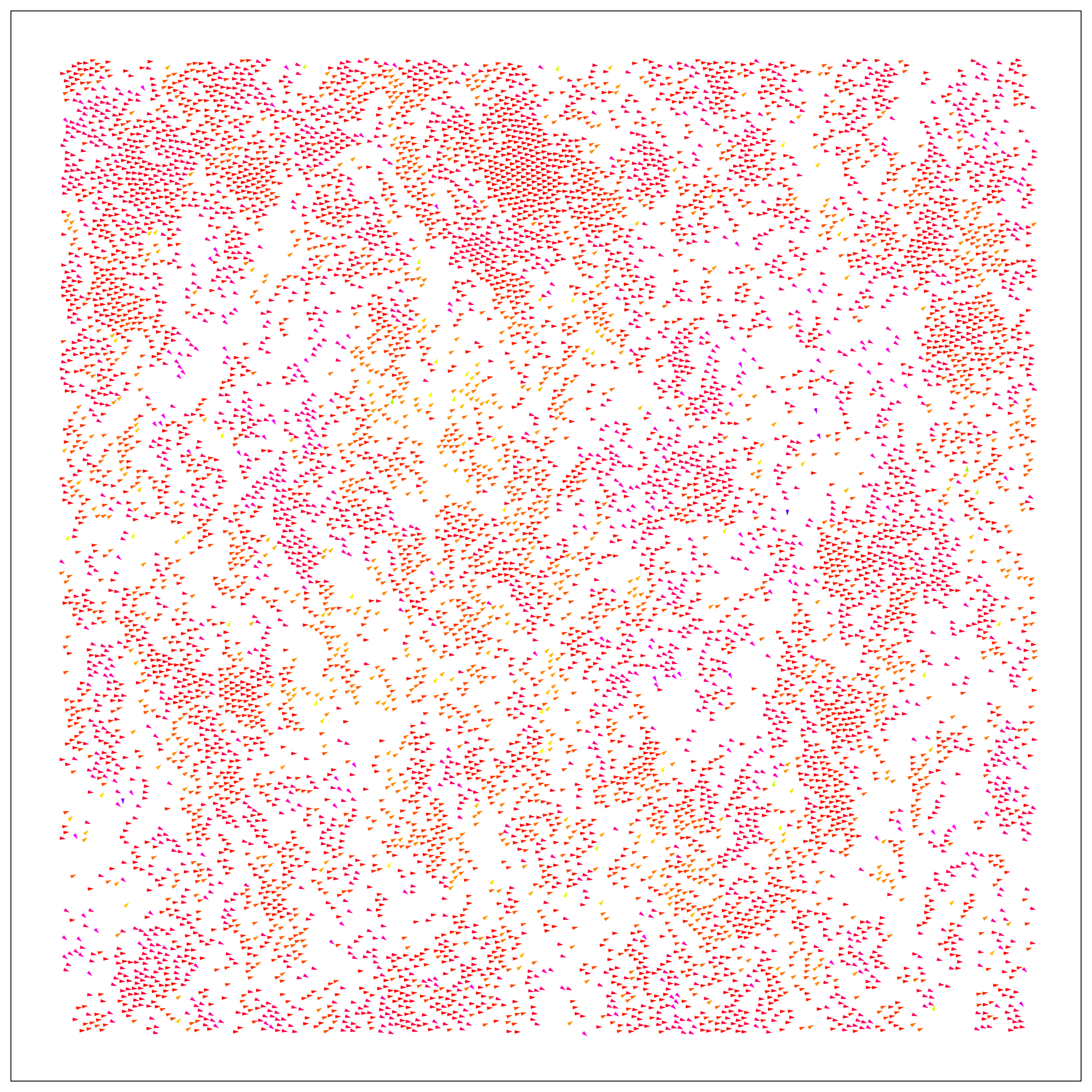}
            \caption{OH-2}
        \end{subfigure}
    \end{tabular}
    \caption{Spatial collective states of systems of self-propelled agents with alignment and repulsion.
    Top: Phase diagram as a function of Peclet number Pe and alignment strength g, showing the fraction of agents in the largest cluster $\Lambda$ (a) and the polarization $\Phi$ (b).
    Bottom: Representative snapshots of the states labeled in panel (a).  
    Disordered (Pe,g)=(32,1) (c); 
    ordered with bands OB (Pe,g)=(32,2) (d); 
    ordered homogeneous OH-1 (Pe,g)=(512,16) (e);
    ordered with clusters OC (Pe,g)=(32,128) (f); 
    and a different ordered homogeneous state OH-2 (Pe,g)=(32,1024) (g).
    Each agent is colored by its heading angle, according to the color disk (top-left inset).}
    \label{fig:snapshots}
\end{figure}

We begin by presenting in Fig.~\ref{fig:snapshots} the phase space of the spatial dynamics.
Panels (a) and (b) display the values of the clustering $\Lambda$ and polarization $\Phi$ for each parameter combination.
Following the work in \cite{zhao2021phase}, we identify five different regions with distinct values of $\Lambda$ and $\Phi$, represented by the characteristic states labeled from (c) to (g) in panel (a).
A snapshot of each corresponding state is presented in panels (c)-(g), where the particles represented by points that are colored according to their heading directions, as indicated in the color disk in the top-left corner.
Panel (c) presents a typical snapshot of the disordered states found in the low $\Phi$ and low $\Lambda$ region of the phase diagram, where there is no orientational order and the particles are homogeneously distributed in space. 
Panels (d) and (f) display the ordered with bands (OB) and ordered with clusters (OC) states, respectively. 
Both have high $\Phi$ and $\Lambda$ values, but OB states form large density bands that are perpendicular to the heading direction, whereas OC states exhibit clusters that are elongated along the heading direction.
Finally, panels (e) and (g) show two different ordered homogeneous states (OH-1 and OH-2), both of which are characterized by high $\Phi$ and low $\Lambda$ values. 
%

\begin{figure}[t!]
    \centering
    \includegraphics[width=\textwidth]{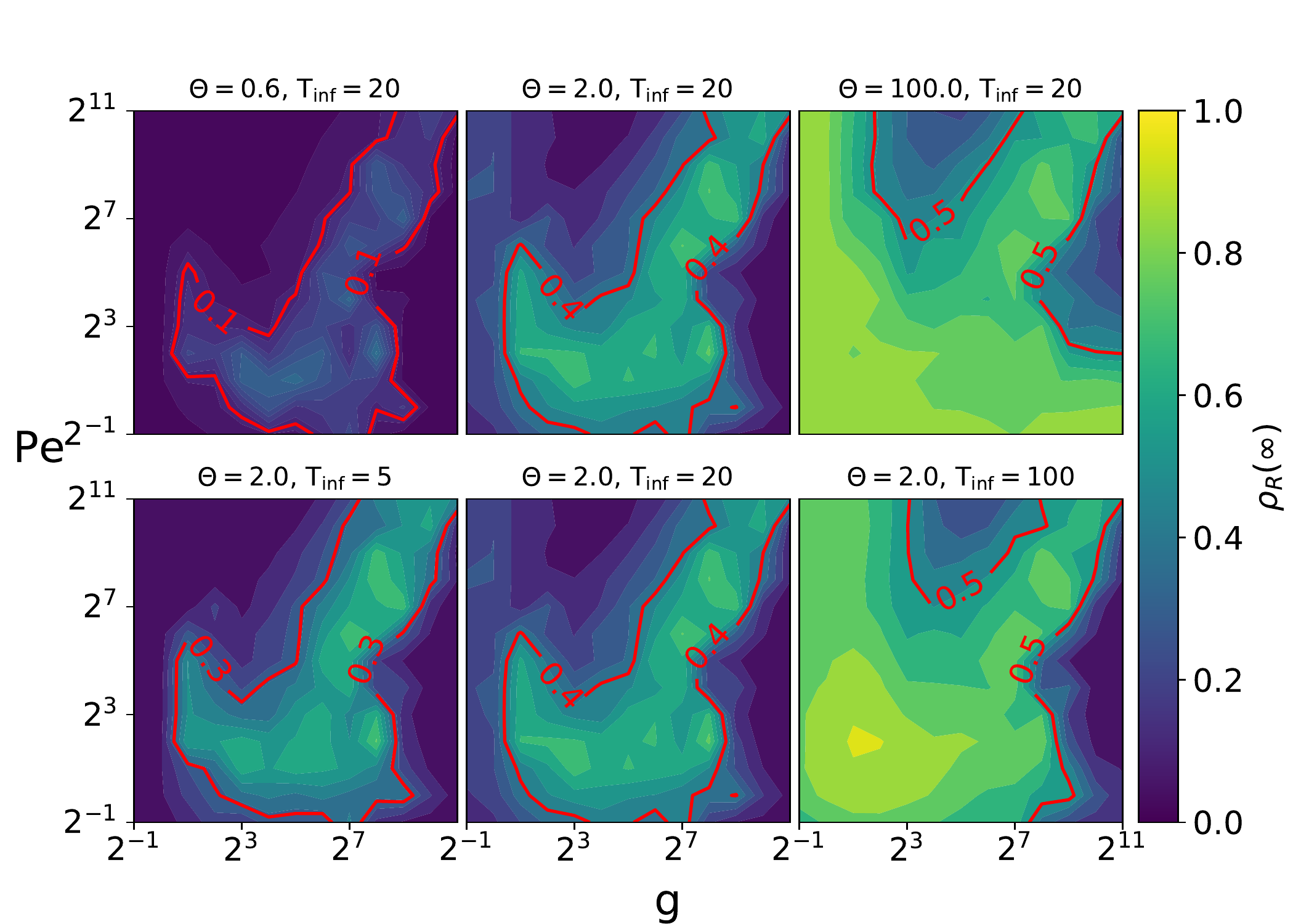}
    \caption{Effectiveness of the epidemic outbreak as a function of the Peclet number Pe and alignment strength g for different values of the infection-lifetime transmissibility $\Theta=\beta_b/\gamma$ and the mean agent infection duration $\mathrm{T_{inf}} =1/\gamma$. 
    The colors represent the fraction of agents $\rho_R(\infty)$ that recovered form the infected state at the end of each run, averaged over 50 simulations. 
    The red isolines highlight the regions with high relative infection levels.
    We observe that the contagion tends to be more effective in the inhomogeneous regions of the phase space, shown on Fig.~\ref{fig:snapshots}, and that it covers a larger fraction of the phase space for higher $\Theta$ or $\mathrm{T_{inf}}$ values.}
    \label{fig:infection-diagram}
\end{figure}

We now examine the effectiveness of the contagion process in different regions of the phase diagram, for various values of the contagion parameters, by computing the fraction of agents that were infected at some point of the epidemic process, which is equal to the final fraction of recovered agents.
The infection spreading will depend on the rate and duration of the contacts between infected and susceptible agents that result from the spatial dynamics, and on the SIR model parameters $\beta_b$ and $\gamma$.
We will express the latter in terms of two quantities that determine the effective contagion rate between susceptible interacting agents: the infection-lifetime transmissibility $\Theta=\beta_b/\gamma$ (a nondimensional number that describes the average total transmission during the infectious period of an agent) and the mean duration of the infectious period of an agent $\mathrm{T_{inf}} =1/\gamma$.
The values of $\Theta$ and $\mathrm{T_{inf}}$ are both characteristics of the disease alone, and do not depend on the contact network. The interplay between these quantities and the spatial dynamics determines the effective infection spread.

Note that there are two limit cases of the spatial dynamics where the final epidemic outcome will only depend on $\Theta$, and not on $\mathrm{T_{inf}}$:
The well-mixed (mean-field) case, where agents are continuously exchanging neighbors, and the static interaction network case (if agents are close enough to form a connected graph), where the neighbors remain fixed.
In both situations, the newly infected agents are continuously interacting with other agents that are potentially susceptible, so the duration $\mathrm{T_{inf}}$ of their infectious state will not affect the outcome. 
Even in these limits, however, the contagion process before reaching this final outcome will also depend on $\mathrm{T_{inf}}$ and on the spatial dynamics of the system, which will affect its timescale \cite{peruani2008dynamics, peruani2019reaction, norambuena2020understanding}.
In more realistic cases where the interactions are determined by complex temporal networks \cite{holme2012temporal} or spatiotemporal dynamics, as in the model considered here, both the epidemic outcome and the contagion process will generally depend on all parameters.

Figure \ref{fig:infection-diagram} presents the effectiveness of the epidemic outbreak in different regions of the phase space for six combinations of the infection parameters.
The final fraction of recovered agents $\rho_R(\infty)$ is displayed as a function of Pe and g, for three $\Theta$ values with fixed $\mathrm{T_{inf}}=20$ (top) and for three $\mathrm{T_{inf}}$ values with fixed $\Theta=2.0$ (bottom).
Note that we repeat the same central panel in the top and bottom rows, for completeness.
We find that, in general, the regions with highest degree of contagion tend to match the regions with highest clustering in Fig.~\ref{fig:clustering-diagram}, and that they grow as $\Theta$ or $\mathrm{T_{inf}}$ are increased. 

The central panels shows that, for an intermediate $\Theta=2.0$ value, the region with high epidemic spreading matches almost exactly the high clustering regions OB and OC in Fig.~\ref{fig:snapshots}, although some simulations with relatively high level of contagion can also be found in the disordered region.
The top right panel shows that, for a very high $\Theta=100$, almost the whole phase space displays high levels of epidemic spreading, with the homogeneously ordered (OH) regions presenting the lowest levels.
The bottom row shows that we obtain equivalent results if we instead vary $\mathrm{T_{inf}}$, but with larger $\rho_R(\infty)$ differences between the regions with high and low epidemic spreading.

\begin{figure}[t!]
    \centering
    \begin{tabular}[b]{cc}
    \begin{subfigure}[b]{0.48\columnwidth}
      \includegraphics[width=1.0\textwidth]{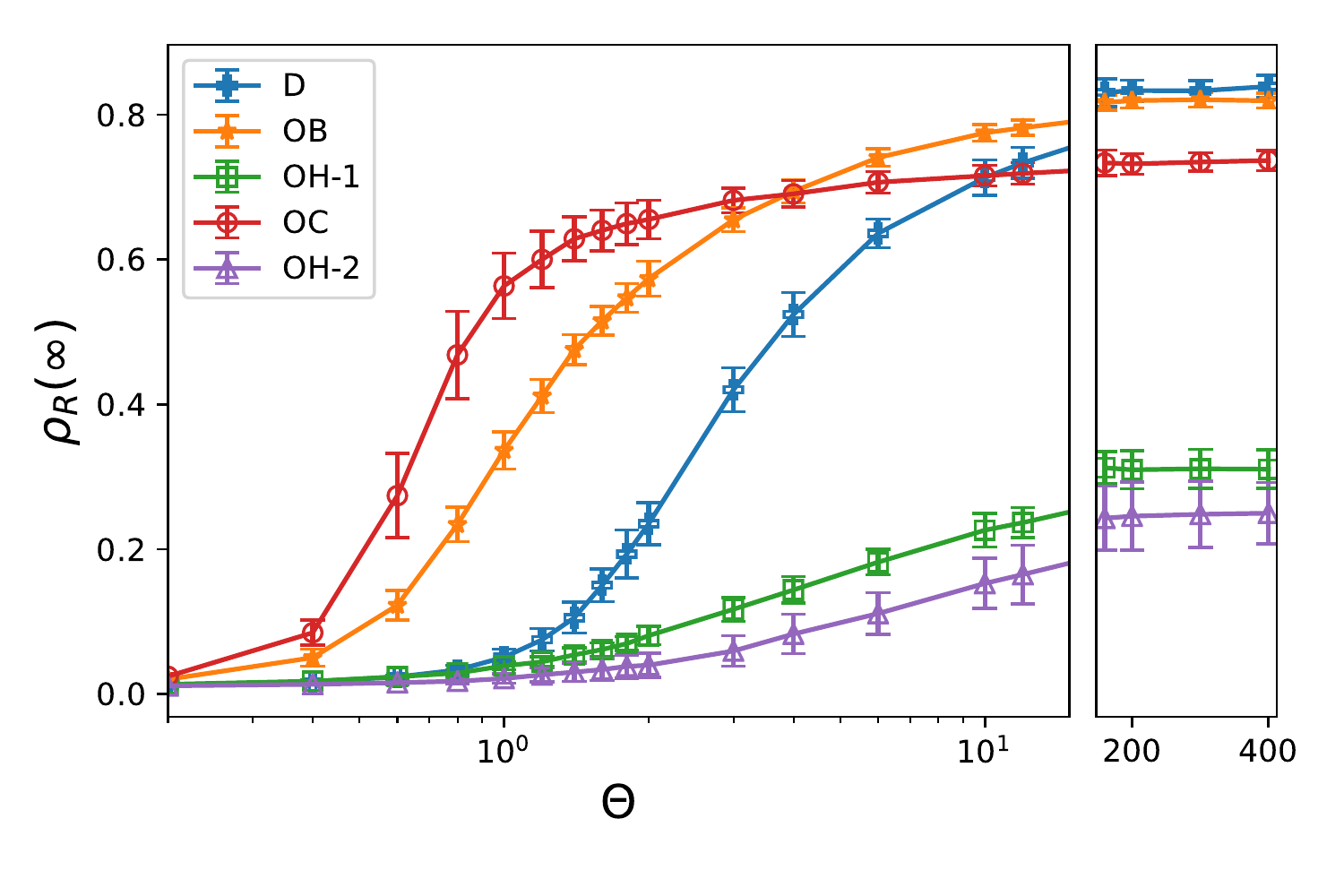}
      \caption{}
      \label{fig:transition-fixed-recoverey}
    \end{subfigure}
    \begin{subfigure}[b]{0.48\columnwidth}
      \includegraphics[width=1.0\textwidth]{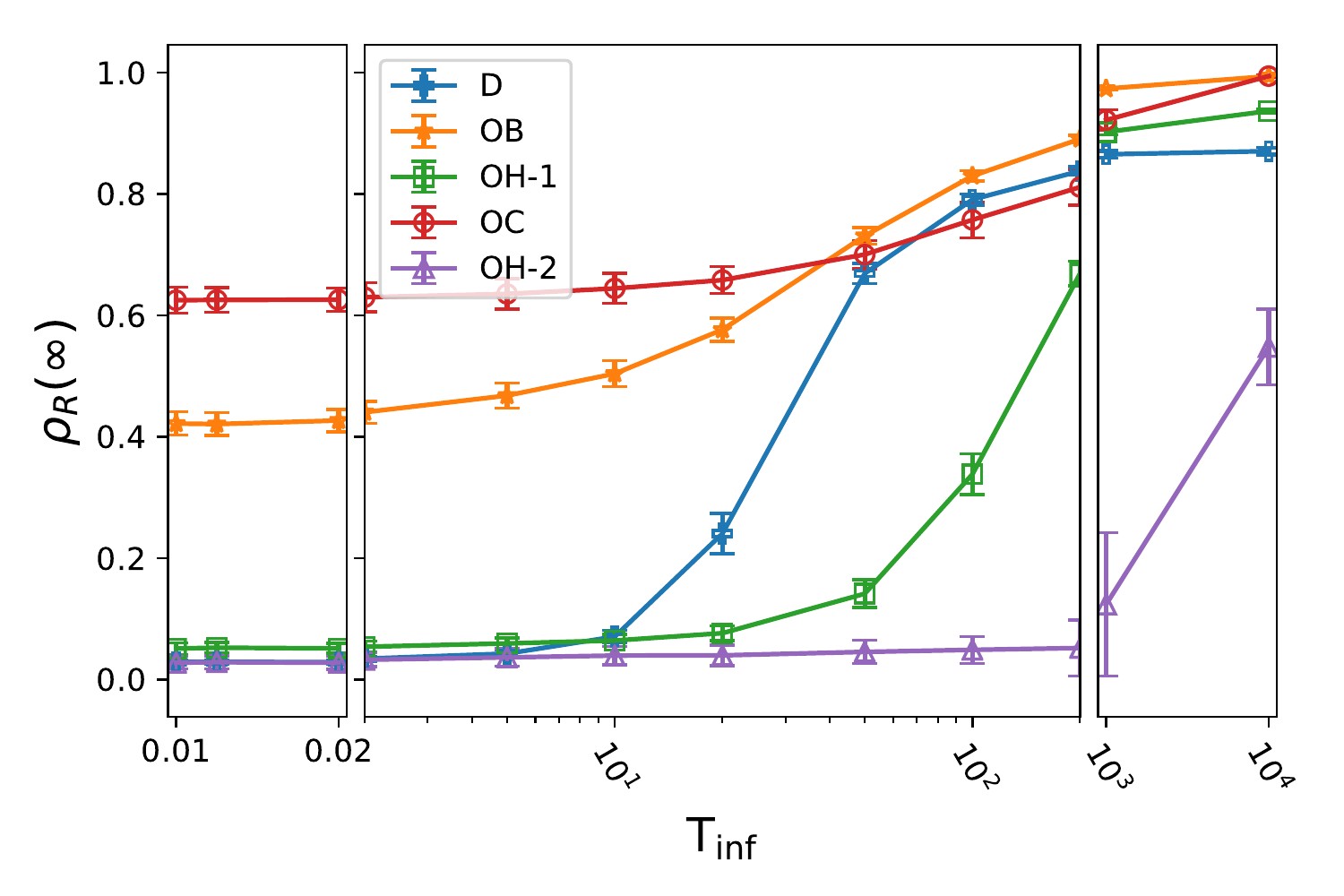}
      \caption{}
      \label{fig:transition-fixed-reproductive}
    \end{subfigure}
    \end{tabular}
    \caption{Effectiveness of the epidemic outbreak as a function of the contagion parameters for the five collective states identified in Fig.~\ref{fig:snapshots}: Disordered (D), ordered with bands (OB), ordered with clusters (OC), and two ordered homogeneous states (OH-1 and OH-2).
    Each panel presents the fraction of agents that were infected and later recovered during the epidemic outbreak $\rho_R(\infty)$ as a function of either the infection-lifetime transmissibility $\Theta=\beta_b/\gamma$ with fixed $\mathrm{T_{inf}}=20$ (a) or of the mean agent infection duration $\mathrm{T_{inf}} =1/\gamma$ with fixed $\Theta=2$ (b).
    Each point shows the mean $\rho_R(\infty)$ value for 50 simulations and the error bars, its standard deviation.}
    \label{fig:infection-transition}
\end{figure}

Figure \ref{fig:infection-transition} provides a more detailed view of the effectiveness of each epidemic outbreak and its onset points, plotting $\rho_R(\infty)$ as a function of $\Theta$ and $\mathrm{T_{inf}}$ for the five collective states identified in Fig.~\ref{fig:snapshots}. Each point displays the mean value obtained for the 50 simulations performed per parameter combination and the error bar shows the corresponding standard deviation.

Panel (a) confirms that outbreak onsets occur at the lowest $\Theta \approx 0.5$ values
in the inhomogeneous (OB, OC) states and at higher $\Theta \approx 1$ values in the homogeneous disordered (D) and homogeneous ordered (OH-1, OH-2) states. The value of $\rho_R(\infty)$ then increases with $\Theta$ at a different rate for each case.
At an intermediate $\Theta \approx 10$, we find that the OB state has the highest mean $\rho_R(\infty)$, followed closely by the OC and disordered states.
For very large $\Theta \gtrsim 200$, the disordered and OB states display the highest, very similar $\rho_R(\infty) \approx 0.8$ values, whereas the OC case seems to saturate at a slightly lower level.
Instead, both OH states display a significantly lower level of epidemic spreading, with a fraction of recovered agents $\rho_R(\infty) \approx 0.3$ that is slightly higher for OH-1 than for OH-2.

Panel (b) shows that the outbreak onsets occur at very different $\mathrm{T_{inf}}$ levels in the disordered, OH-1, and OH-2 states (order $10^1$, $10^2$, and $10^3$, respectively). 
These three homogeneous states eventually reach high infection levels, but only for very high $\mathrm{T_{inf}}$ values.
This panel also shows that, in the $\Theta=2$ case considered here, states OC and OB develop high levels of infection for all $\mathrm{T_{inf}}$ values.
Even at a very low $\mathrm{T_{inf}}=10^{-2}$ we find that, remarkably, they still display significant outbreaks, with $\rho_R(\infty) \gtrsim 0.4$.
In the opposite limit, for a very high $\mathrm{T_{inf}}=10^{4}$, both converge to a similar state of full epidemic spreading, with $\rho_{R}(\infty)\approx 1$.

In the following section, we will discuss the underlying mechanisms that lead to the significant variations in the final level of infection displayed above by the different states of collective dynamics. 

\section{Discussion}

The results in the previous sections show that the infection parameters and the spatial dynamics both play an essential role in the contagion spreading.
In particular, we observe that complex self-organized states tend to result in higher levels of infection than the more commonly studied states with either well-mixed or static interactions. Furthermore, the specific nature of the spatial dynamics appears to play an essential role that cannot be captured by standard averaged network-based descriptions.
We will analyze here in detail the spreading process in the different identified collective regimes, as a function of the infection parameters $\Theta$ and $\mathrm{T_{inf}}$.

We begin by considering the simplest case, the homogeneous disordered state, in which all agents are continuously mixing with similar contact dynamics that do not change over time.
The contagion process is the closest here to a mean-field SIR system, where the final epidemic spreading can be predicted if $\mathrm{T_{inf}}$ is sufficiently large \cite{peruani2019reaction}.
If this is the case, a spatially localized infection will always continue to spread, so we can estimate through a semi-analytical mean-field calculation the fraction of agents that will be infected in the wake of the front.
We present these results in the Supplementary Material, which we obtained by extending similar analyses performed in 
\cite{peruani2008dynamics,peruani2019reaction,norambuena2020understanding} 
for the SIS infection model spreading through active Brownian particles.
For a small $\mathrm{T_{inf}}$, however, the mixing timescale can be comparatively too long for new susceptible agents to continue to be exposed to infected agents before they recover, since the front advances slower over time, and the infection will eventually die out.
The lack of mixing due to short $\mathrm{T_{inf}}$ can be overcome by increasing the spatial density of initially infected agents, as we also discuss in the Supplementary Material. In a scenario, with a spatially homogeneous distribution of initial infections (or infection clusters), we may observe a percolation-like merging of the different locally expanding fronts, which could eventually result in almost the entire system being infected even for short $\mathrm{T_{inf}}$. 

In the ordered homogeneous states, OH-1 and OH-2, the high level of polarization results in slow mixing and thus low contact rate between new neighbors. Therefore, sufficiently high $\Theta$ and $\mathrm{T_{inf}}$ values are crucial for the contagion process to develop. 
For low $\mathrm{T_{inf}}$, the interaction network will not experience significant changes before the infected individuals recover, and can thus be approximated by a static network. 
In this case, the epidemic spreading can be viewed as a percolation process on this network. For a low mean density, the sparse connectivity can therefore place the system below the percolation threshold, stopping contagion even for extremely high $\Theta$ values, as shown in Fig.~\ref{fig:transition-fixed-recoverey}.
For higher $\mathrm{T_{inf}}$, local velocity fluctuations will lead to the accumulation of contacts with different neighbors during this period, increasing the effective connectivity, which is now given by the time-averaged interaction network. This can result in higher epidemic spreading, as shown in Fig.~\ref{fig:transition-fixed-reproductive}. 
Note, however, that these fluctuations can also produce more mixing between susceptible and recovered agents at the infection propagation front, which would reduce the fraction of agents that are available there for contagion and thus limit the epidemic growth.
This effect was observed in \cite{rodriguez2019particle} for a system of particles moving in straight lines with random directions, where an optimal mixing speed that minimizes spreading was identified in a regime between quasi-static network percolation and high mixing.
Although we did not detect this type of effect in our simulations, it could be present for other parameter combinations.

\begin{figure}[t!]
    \centering
    \includegraphics[width=\textwidth]{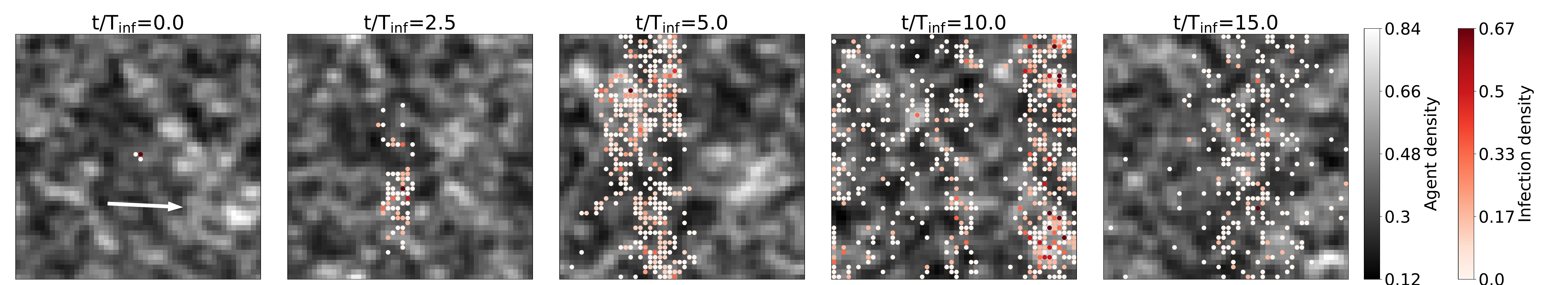}
    \caption{Formation of the initial infection band in an ordered homogeneous state. The greyscale bins show the smoothed local mean density of all agents within their square area; the overlaying white-to-red scale dots, the corresponding density of infected agents. 
    The panels display the temporal evolution of a single infection process (with $\mathrm{T_{inf}}=10^3$ and $\Theta=2$) started by 10 agents at the center of the arena. The underlying spatial dynamics corresponds to the ordered homogeneous state OH-1 presented in Fig.\ref{fig:snapshots}, with (Pe,g)=(512,16), which is advancing to the right as indicated by the white arrow.}
    \label{fig:fields}
\end{figure}

A more detailed inspection of the contagion process in the ordered homogeneous states shows that it develops spatial dynamics that cannot be described by simple percolation or mixing approximations. 
In particular, in this regime the agents are known to display superdiffusive displacements perpendicular to the mean heading \cite{toner1998flocks, toner2005hydrodynamics}, which will result in a faster propagation of the infection along this direction.
Starting from a localized outbreak, we thus observe the rapid initial formation of a band-like infected region perpendicular to the mean agent orientation, which is then advected by the mean flow, as shown in Fig.\ref{fig:fields}. 
This type of contagion process was identified in all our simulations of states OH-1 and OH-2; we expect it to also be present in any ordered homogeneous state of self-propelled particles with similar dynamics.

Although the two ordered homogeneous states seem to display a similar contagion process, the level of epidemic spreading is strongly conditioned by their specific spatial dynamics.
Indeed, since OH-2 has a higher alignment strength $g$ than OH-1, it will present a smaller contact rate and longer typical contact duration \cite{zhao2021phase}, which will result in a much lower effective contagion rate. This explains its higher onset value ($\mathrm{T_{inf}} \gtrsim 10^3$) and the fact that, even at $\mathrm{T_{inf}} = 10^4$, the effectiveness of the epidemic outbreak only reaches $\rho_R(\infty) \approx 0.5$ in Fig.~\ref{fig:transition-fixed-reproductive}.

The details of the infection dynamics also appear to be different, as shown in the simulation videos included in the Supplementary Material.
The outbreaks in state OH-1 typically develop approximately homogeneously throughout the simulation arena. By contrast, the outbreaks in state OH-2 appear to develop mainly inside large clusters that emerge sporadically and survive for relatively short periods of time during the simulations, spreading less effectively outside of them.
We note, however, that we only studied the OH-2 outbreaks for $\mathrm{T_{inf}}=10^4$, where they become more prevalent. 
We did not explore the asymptotic behavior in this state for larger $\mathrm{T_{inf}} \gg 10^4$ because this would require extremely long simulations that go beyond the scope of this work, given that the time needed to reach a final infection state grows with $\mathrm{T_{inf}}$ and that the dynamics for high coupling strengths can only be resolved with very small timesteps.
Finally, we point out that the differences in the contagion processes of OH-1 and OH-2 are expected to disappear for large $\mathrm{T_{inf}}\to\infty$, since the infection timescales will become so long that the spatial dynamics will effectively result in well-mixed systems over comparatively short timescales, and can thus be averaged over time.

In the ordered states with bands, or OB states, the system can be divided into two regions: the high density, high polarization regions within the bands, and the low density, low polarization, gas-like regions outside of the bands.
The infections that develop outside of the bands will thus take place in regions that are similar to the disordered states and have lower density than the system average.
Therefore, for the contagion to spread in this region, $\Theta$ and $\mathrm{T_{inf}}$ must be higher than the corresponding epidemic thresholds in the disordered state, since lower density implies less interactions between infected and susceptible agents. Otherwise, our simulations show that the infection will die out quickly, unless it is transmitted to a band.
Instead, the infections that develop within the bands take place in regions with higher local density and polarization. Agents will have more interactions that last longer while the combination of high density and repulsion will inhibit mixing, so the contagion process will be predominantly driven by percolation.
Our simulations show that successful infection propagation within the bands requires a significantly lower $\Theta$ and $\mathrm{T_{inf}}$.

In this context, in the OB states the contagion spreading process near the epidemic threshold will typically develop as follows.
The initial infection propagates more efficiently within the bands, due to their higher local densities.
Then, given that bands advance faster than the mean agent velocity between bands, due to their higher polarization, once a significant fraction of a band is infected, it will develop a contagion process that can span the system by sweeping through the arena, reaching the susceptible agents in the gas-like state outside the band.
Therefore, the epidemic spreading is initially controlled by $\Theta$, which needs to be high enough to grow the infection within a band, and later by $\mathrm{T_{inf}}$, which will determine the level of contagion outside the band when we compare it to the time required for the band to swipe the system.

The ordered clustered states, or OC states, typically contain one or more large polarized clusters that are elongated along their heading direction, surrounded by lower density regions.
As in the OB states, the contagion develops differently inside and outside of these clusters. In contrast to the OB states, however, the low density regions outside of the clusters have relatively high levels of polarization, albeit smaller than within clusters.
The contagion dynamics in these regions is thus similar to the percolation process in the homogeneous ordered states, but will require higher $\Theta$ and $\mathrm{T_{inf}}$ values to propagate successfully, due to their lower density. 
Otherwise, our simulations show that any infection that starts outside the clusters will die out quickly, unless it spreads into a cluster.
We also observed in our simulations that, when contagion does develop outside of the clusters (for higher $\Theta$ or $\mathrm{T_{inf}}$ values), it will spread faster perpendicular to the heading direction, as in the OH states.
Inside the clusters, the contagion spreads primarily through percolation on a quasistatic network, due to the weak local mixing.
Here the densities are higher than in the homogeneous ordered states and than inside the bands in the OB states, so the infection process is very effective and requires the lowest $\Theta$ and $\mathrm{T_{inf}}$ values to spread. 

After a cluster gets infected in an OC state near the epidemic threshold, the contagion process to other regions can be quite complex, as it involves two opposing mechanisms.
On the one hand, clusters do not frequently interact with new off-cluster agents or with other clusters, since all the particles in the system are approximately aligned. This implies that, for low or intermediate $\mathrm{T_{inf}}$, the infection inside a cluster may never spread outside. 
On the other hand, clusters exhibit rich fission-fusion dynamics through which an infected cluster that disaggregates can infect other clusters, as its resulting fragments aggregate into them (see videos in Supplementary Material).
As a result, although the infection within a cluster can be very effective even for low $\Theta$ and $\mathrm{T_{inf}}$, its spreading to the rest of the system depends on the balance between these two processes that develops in each realization of the spatial dynamics.
It is this lack of a reliable mechanism for spreading the infection throughout the system that makes the OC states reach a slightly lower $\rho_R(\infty)$ level than the disordered and OB states for intermediate and high $\Theta$ and $\mathrm{T_{inf}}$ values in Figs.~\ref{fig:infection-diagram} and \ref{fig:infection-transition}.
Indeed, in these cases, it is the corresponding mixing and swiping by the bands that ensures the global spreading of the infection when the system is above the epidemic threshold.

The analyses carried out above can be summarized in terms of contagion processes that develop on evolving networks.
In this context, the infection-lifetime transmissibility $\Theta$ becomes a determinant contagion parameter that, if the network connectivity dynamics is well-described by mean contact and duration rates, can be directly linked to what is know as the basic reproduction number in standard epidemiology.
This quantity, often labeled $R_0$, is defined as the mean number of infections directly generated by one infected agent in a susceptible population.
When the temporal dynamics of the network topology cannot be well described by averages, however, the infection duration $\mathrm{T_{inf}}$ also plays an important role by setting the timescale of the contagion process relative to the network rewiring dynamics.
As discussed above in terms of the spatial dynamics, for small $\mathrm{T_{inf}}$ values, the rewiring timescale is comparatively long and the contagion is determined by percolation on a static network. For the density considered here, all the homogeneous states (disordered, OH-1, and OH-2) are below the percolation transition and no macroscopic outbreaks are observed.
For large $\mathrm{T_{inf}}$ values, the epidemic outbreak is determined by the properties of the time-aggregated network and can fall into one of two categories: 1) ordered states where the nearest neighbors (within or beyond the interaction range) remain relatively unchanged and the contagion is driven by an expanding front, and 2) disordered states where agents are continuously interacting with new neighbors and the contagion is driven by mixing.
For intermediate $\mathrm{T_{inf}}$ values, the properties of the spatiotemporal structures that emerge through self-organization play a crucial role in the infection process.
The epidemic outcome will thus also depend on the ratio between the contagion timescale $\mathrm{T_{inf}}$ and the timescale governing the evolution of these structures.
In the OB and OC states, for example, we find that $\Theta$ controls whether a contagion outbreak happens, whereas $\mathrm{T_{inf}}$ determines the total scope of the outbreak, as shown in Fig.~\ref{fig:infection-transition}.
We expect this situation to change, however, if the value of the infection duration $\mathrm{T_{inf}}$ is very high, which will make all the timescales of the spatial dynamics negligible, resulting in an effective contagion process that is equivalent to the well-mixed case.

\section{Conclusions \& Broader Implications}

In this work, we investigated a simple SIR epidemic process in a system of self-propelled agents with alignment interactions. We showed that the contagion dynamics and the overall outbreak magnitude depend strongly on the spatiotemporal collective states. In particular, we found that effective infection spreading can be favored by self-organization.
Our analyses followed a minimal modeling approach to show potential generic connections between emergent spatiotemporal dynamics and the spreading of epidemics that could have strong impact on a variety of real-world scenarios.
Although our results were mainly expressed in terms of epidemic spreading, we note that they can also describe other processes, such as the propagation of different types of information within groups of agents.

Revealing the impact of self-organized collective behavior \cite{zhao2021phase,klamser2021evolution} on contagion processes is essential for understanding the functional aspects of a variety of behaviors exhibited by biological collectives (such as the risk-related behavior described in \cite{sosna2019individual}).
In bacterial swarms, for example, the spreading of genetic information (through direct contacts or via bacteriophages \cite{davison1999genetic}) is conditioned by the self-organized coordinated motion and cluster formation that emerges from simple physical interactions \cite{peruani2012collective,chen2012scale,be2020phase}.
These genetic propagation dynamics are believed to play an important role in key processes such as the spread of antibiotic resistance in bacterial populations \cite{mazaheri2011bacteriophage,muniesa2013potential}.
A better understanding of this type of phenomena will thus require further studies of the interplay between infection spreading and collective spatiotemporal dynamics that can be built on the results obtained in our work.

At the macroscopic agent scale, the relationship that we analyzed here between spatiotemporal self-organization and contagion dynamics is expected to be highly relevant for a variety of systems, ranging from animal groups to human crowds, which are known to often form dense gatherings with complex fission-fusion dynamics (see, e.g., \cite{sundaresan2007network,silk2014importance,bierbach2020interaction}). 
This could affect processes that require containment, such as disease spreading, or that are typically beneficial, such as the diffusion of information. 

Despite using highly idealized models, our work could have relevant implications in the development of strategies for the control of epidemics.
For example, our results show that, for a broad range of parameters, outbreaks are significantly reduced in the homogeneous ordered states when compared to other states with the same, relatively low, mean density. 
In the context of the current COVID-19 pandemic, this suggests the development of crowd control strategies that favor the homogeneous spreading of individuals to help minimize contagion during large gatherings.
On a more general note, our minimal agent-based descriptions show that the contagion dynamics is often not adequately captured by the averaged network approaches commonly used to predict epidemic spreading.
This is especially true in the presence of self-organized spatiotemporal features, which could help explain the very limited success that has been achieved in predicting outbreaks during this pandemic.

Future studies could improve the level of description reached by the type of minimal models considered in this paper by including other potential couplings between the spatial dynamics and the contagion process. In a recent study \cite{levis2020flocking}, for example, flocking agents alter their motion when they get infected, which can strongly affect the contagion spreading. This type of additional considerations could help complete the characterization of simple systems that combine spatial and epidemic models, and will be left for future work.

\section*{Acknowledgement}

YZ is grateful for the financial support from the China Scholarship Council (CSC).
PR and YZ acknowledge funding by the Deutsche Forschungsgemeinschaft (DFG, German Research Foundation) through the Emmy Noether Programm - RO 4766/2-1 and under Germany’s Excellence Strategy – EXC 2002/1 “Science of Intelligence” – project number 390523135.
%
The work of CH was partially funded by CHuepe Labs Inc.

\newcommand{\newblock}{}
\bibliographystyle{unsrt}
\bibliography{references}
\end{document}